# Radio Frequency Phototube, Optical Clock and Precise Measurements in Nuclear Physics

*Amur Margaryan*
*Yerevan Physics Institute after A.I. Alikhanian*
*Yerevan, Armenia*

## Abstract

Recently a new experimental program of novel systematic studies of light hypernuclei using pionic decay was established at JLab (Study of Light Hypernuclei by Pionic Decay at JLab, JLab Experiment PR-08-012). The highlights of the proposed program include high precision measurements of binding energies of hypernuclei by using a high resolution pion spectrometer, HπS. The average values of binding energies will be determined within an accuracy of ~10 keV or better. Therefore, the crucial point of this program is an absolute calibration of the HπS with accuracy $10^{-4}$ or better.

The merging of continuous wave laser-based precision optical-frequency metrology with mode-locked ultrafast lasers has led to precision control of the visible frequency spectrum produced by mode-locked lasers. Such a phase-controlled mode-locked laser forms the foundation of an optical clock or "femtosecond optical frequency comb (OFC) generator," with a regular comb of sharp lines with well defined frequencies. Combination of this technique with a recently developed radio frequency (RF) phototube results in a new tool for precision time measurement. We are proposing a new time-of-flight (TOF) system based on an RF phototube and OFC technique. The proposed TOF system achieves 10 fs instability level and opens new possibilities for precise measurements in nuclear physics such as an absolute calibration of magnetic spectrometers within accuracy $10^{-4}$ - $10^{-5}$.

## 1. Introduction

The binding energies of the $\Lambda$ particle in the nuclear ground state give one of the basic pieces of information on the $\Lambda$-nucleus interaction. Most of the observed hypernuclear weak decays take place from the ground states, because the electromagnetic interactions or Auger neutron emission process are generally faster than the weak decay of the $\Lambda$ particle. The binding energy of $\Lambda$ in the ground state is defined by:

$$B_\Lambda(g.s.) = M_{core} + M_\Lambda - M_{HY}.$$

The mass $M_{core}$ is merely the mass of the nucleus that is left in the ground state after the $\Lambda$ particle is removed. The binding energies $B_\Lambda$ have been measured in emulsion for a wide range of light ($3 \leq A \leq 15$) hypernuclei [1]. These have been made exclusively from weak $\pi^-$-mesonic decays.

From precise hypernuclear ground state binding energies and detailed hypernuclear low level structure, we can establish the $\Lambda N$ spin-dependent (spin-spin, spin-orbit, and tensor forces) interaction strengths, and then investigate the $\Sigma N - \Lambda N$ coupling force, and charge symmetry breaking. Experimental information on these characteristics of the $\Lambda N$ interaction plays an essential role to discriminate and improve baryon-baryon interaction models, not only those based on the meson-exchange picture but also those including the quark-gluon degree of freedom, toward unified understanding of the baryon-baryon interactions. In addition, understanding of the YN and YY interactions is necessary to describe high density nuclear matter containing hyperons.

The binding energies of light hypernuclei $^3_\Lambda H, ^4_\Lambda H$, and $^4_\Lambda He$ are the most valuable experimental information for checking different models of YN interaction. In the Table 1 taken from reference [2] lists the results of the $\Lambda$ separation energies obtained as a result of ab initio calculations using YN interactions with an explicit $\Sigma$ admixture. It is demonstrated that for future theoretical developments more precise experimental measurements for binding energies are needed.

Table 1: $\Lambda$ separation energies, given in units of MeV, of A = 3-5 $\Lambda$ hypernuclei for different YN interactions.

| YN | $B_\Lambda(^3_\Lambda H)$ | $B_\Lambda(^4_\Lambda H)$ | $B_\Lambda(^4_\Lambda H^*)$ | $B_\Lambda(^4_\Lambda He)$ | $B_\Lambda(^4_\Lambda He^*)$ | $B_\Lambda(^5_\Lambda He)$ |
|---|---|---|---|---|---|---|
| SC97d(S) | 0.01 | 1.67 | 1.2 | 1.62 | 1.17 | 3.17 |
| SC97e(S) | 0.10 | 2.06 | 0.92 | 2.02 | 0.90 | 2.75 |
| SC97f(S) | 0.18 | 2.16 | 0.63 | 2.11 | 0.62 | 2.10 |
| SC89(S) | 0.37 | 2.55 | Unbound | 2.47 | Unbound | 0.35 |
| Experiment | $0.13 \pm 0.05$ | $2.04 \pm 0.04$ | $1.00 \pm 0.04$ | $2.39 \pm 0.03$ | $1.24 \pm 0.04$ | $3.12 \pm 0.02$ |

The existing situation can be summarized by the help of words of R. Dalitz [3]:
"$^3_\Lambda H$ was well known very early and has been studied a great deal. Its $B_\Lambda$ value is quite small and difficult to measure. It was the first hypernucleus to be considered a "$\Lambda$ halo". The value of $0.13 \pm 0.05$ MeV by Don Devis [1] quoted above was from emulsion studies. From HeBC studies, Keyes et al. [4] have given $0.25 \pm 0.31$ MeV for all events ($^3He\pi^-$) but got $-0.07 \pm 0.27$ when they added in all other $\pi^-$ modes, which is not reassuring. For $R_3 = n(^3He\pi^-)/n(^3_\Lambda H \to all\pi^- \bmod es)$ they give $R_3 = 0.36 \pm 0.07$, and consider this to correspond to $0.11^{+0.06}_{-0.03}$ MeV for its $B_\Lambda$ value. I feel that we are far from seeing the end of this road. A good deal of theoretical work on this 3-body system would still be well justified."

A new experimental program for precise measurement of binding energies, $B_\Lambda$, for a light (A $\leq$ 12) mass range of hypernuclei have been established at CEBAF, by using again $\pi^-$ -mesonic decays [5]. It was proposed to measure the momentum of decay $\pi^-$ -mesons by using a high resolution pion spectrometer, H$\pi$S. The expected resolution is about 100 keV, which is 5-10 times better than in the case of emulsion. Average values of the $B_\Lambda$ will be determined within an error of about 10 keV or better and absolute calibration of the H$\pi$S within accuracy 10 keV or better is a crucial condition for the PR-08-012 experiment.

The recent development of ultrabroad femtosecond optical frequency combs (OFC) based on mode-locked lasers has provided a relatively simple and straightforward way to translate optical frequency standards to other optical or microwave frequencies [6, 7]. We are proposing a new time-of-flight, TOF system based on a recently developed radio frequency (RF) phototube [8] and OFC technique. In this approach the microwave output from the OFC synthesizer is used for driving the RF phototube, which detects simultaneously elementary particles and light bunches of OFC. The proposed new method allows avoiding of excess photodetection noise [9] and internal drift of the tube [10] and achieves the 10 fs instability level (to be compared to ~1 ps, available with the regular timing technique). The proposed TOF system opens new possibilities for precise measurements in nuclear physics such as absolute calibration of the magnetic spectrometers within accuracy $10^{-4}$ - $10^{-5}$.

## 2. The Optical Clock

All clocks consist of two major components: some device that produces periodic events or "clock ticks," and some means for counting, accumulating, and displaying each tick. For example, the swing of a pendulum provides the periodic events that are counted, accumulated, and displayed by means of a set of gears driving a pair of clock hands. If the frequency of the oscillator is locked to the atomic transition frequency, then the time generated can have improved long-term stability and accuracy. For an atomic clock based on a microwave transition, high-speed electronics count and accumulate a defined number of cycles of the reference oscillator to mark a second of time. The basic concepts are the same for an atomic clock based on an optical transition at a much higher frequency.

The recent development of ultrabroad femtosecond optical combs based on mode-locked lasers has provided a relatively simple and straightforward way to translate optical frequency standards to other optical and microwave frequencies [6, 7]. The Fourier transform of a femtosecond pulse train is a comb of evenly spaced frequencies, typically spanning some 10 % of the optical range. The frequency of individual femtosecond comb teeth, $f_n$, can be uniquely described by the relation $f_n = n \times f_r + f_0$, where $n$ is a large integer (~ 500,000), $f_r$ is the repetition rate of the laser, and $f_0$ is the offset of the entire comb from harmonics of $f_r$. Controlling two degrees of freedom of the comb precisely determines the frequency of each comb tooth $f_n$. Both $f_r$ and $f_0$ are microwave frequencies, whereas the comb teeth $f_n$ are optical frequencies; therefore the femtosecond comb architecture provides a phase-coherent translation between the two frequency ranges.

There are different schemes possible to achieve stabilization of the comb, each allowing examination of different quantities. One specific use of the femtosecond comb synthesizer is that of a microwave synthesizer. In this case, the comb is stabilized to an optical oscillator and the microwave output of the comb appears as the repetition rate of the laser. The quality of the stabilized optical pulse trains can be evaluated with the basic apparatus based on two different approaches. The first approach is an optical cross-correlation technique; the second one is a classic heterodyning of two RF frequencies synthesized by using fast photodiodes. It was demonstrated that the fractional frequency instability of the femtosecond comb microwave synthesizer is $2 \times 10^{-14} / \tau$, and could be improved by at least a factor of 10 upon elimination of excess photodetection noise [9].

## 3. The RF Phototube
### 3.1 Principles of operation

The operational principles of the technique are described in Ref. [8] and the schematic layout of the RF phototube is shown in Fig. 1. The primary photon pulse hits the photocathode (1) and produces photo-electrons (PEs). These electrons are accelerated by a voltage $V$ applied between the photocathode and an electron transparent electrode (2). Due to this the time spread of electrons is minimized. The electrostatic lens (3) then focuses and transposes the electrons isochronously onto the screen (7) at the far end of the tube, where the secondary electron (SE) detector is placed. The time structure of the produced PE bunch is identical to that of the light pulse. Along the way the electrons are passed through the circular sweep RF deflection system, consisting of electrodes (4) and the $\lambda/4$ coaxial RF cavity (6), which operates at 500 MHz. They are deflected and form a circle on the screen of the detector, where the time structure of the input photon signal is transferred into spatial PE image (5) on a circle, calibrated in time by the sweep itself, and detected. The detection of the RF analyzed PEs is accomplished with a position sensitive detector based on multichannel plates. In this way the RF phototube transposes the linear time axis into a circle one. The corresponding phases of the applied RF field then are fixed, detected and stored. Meanwhile in such a detection system the timing error sources are minimized, because PEs is timed before the necessary further signal amplification and processing. Internal resolution of such a system is about 2 ps. The RF deflection system operates as a high frequency oscilloscope and times PEs before the electron multiplication process starts in the detector. The electron multiplication process in this case is used to determine the position on the scanning circle.

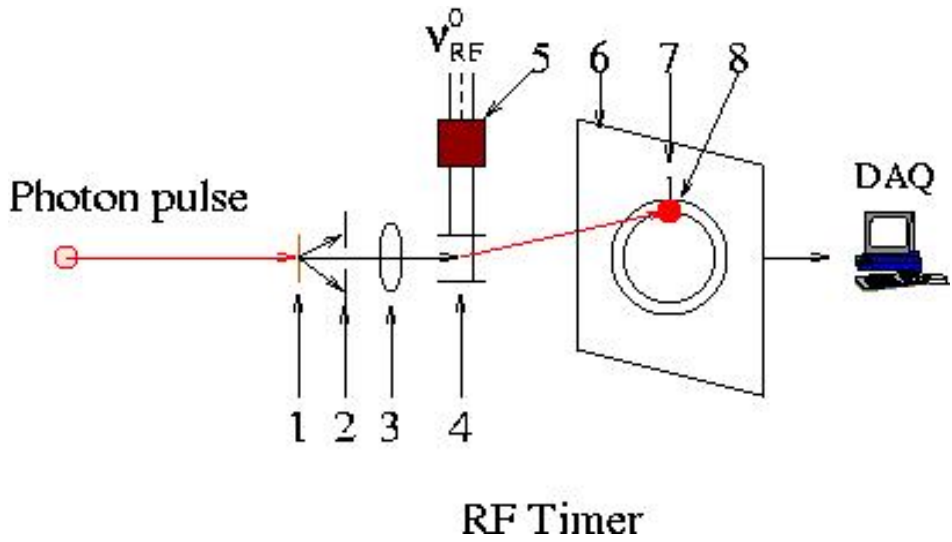

Figure 1: Schematic layout of the RF phototube. 1- photo-cathode, 2- electron transparent electrode, 3- electrostatic lens, 4- RF deflector, 5- $\lambda/4$ coaxial RF cavity, 6- SE detector, 7- arbitrary reference, 8- image of photo-electrons.

Two different ways of recording the PEs images can be implemented: either a "single sweep" can give the required data or an averaged set of data can be recorded in the so-called "synchroscan mode."

### 3.2 Stand-alone operation: random photon source

The time or phase instabilities of the RF phototube are conditioned by instabilities of the RF oscillator and phototube.

The signal $V_{RF}$ delivered by an RF oscillator can be written [11]

$$V_{RF}(t) = V_{RF}^0 \sin[2\pi \nu_{RF}^0 t + \phi_{RF}(t) + \phi_{RF}^0],$$

where we have assumed that the amplitude $V_{RF}^0$ is constant. The quantity $\nu_{RF}^0$ is the constant nominal frequency and refers to the definition of the second, and $t$ is the ideal proper time. The phase $\phi_{RF}(t)$ contains the deviations, both random and systematic, relative to the ideal periodic variations, $\nu_{RF}^0$. The quantity $\phi_{RF}^0$ represents the nominal phase.

The instantaneous frequency $\nu_{RF}(t)$ is defined by

$$\nu_{RF}(t) = \nu_{RF}^0 + \frac{1}{2\pi}\frac{d\phi_{RF}(t)}{dt}.$$

The relative frequency offset of the oscillator is given by

$$y_{RF}(t) = \frac{\nu_{RF}(t)}{\nu_{RF}^0} - 1.$$

It represents as a relative value the instantaneous departure of the oscillating frequency from its minimal value. This is very small for the highly stable oscillators.

The clock reading is incremented by the nominal value $1/\nu_{RF}^0$ of the period every time the total phase $2\pi\nu_{RF}^0 t + \phi_{RF}(t) + \phi_{RF}^0$ increases by $2\pi$. The time indicated since the time origin is therefore

$$t_{RF} = t + x_{RF}(t),$$

where we have set

$$x_{RF}(t) = \frac{\phi_{RF}(t)}{2\pi v_{RF}^0}.$$

The quantity $t_{RF}$ is the proper time of the RF oscillator. The proper time deviation resulting from imperfection in the oscillator is thus equal to $x_{RF}(t)$. This is the time instability of the clock with respect to an ideal reference time scale.

From the above definition, it is easy to show that

$$y(t) = \frac{dx(t)}{dt},$$

here- $y(t)$ represents the relative frequency offset of an oscillator or a clock as a function of $t$ and $x(t)$ its proper time deviation.

The photoelectron passes through the RF deflector at the time moment $t^i$ and fixes the total phase of the RF oscillator:

$$\Phi_{RF}^i = 2\pi v_{RF}^0 t^i + \phi_{RF}(t^i) + \phi_{RF}^T(t^i) + \phi_{RF}^{0,T},$$

somewhere on the scanning circle, determined by $\phi_{RF}^{0,T}$. For the arbitrarily calibrated position sensitive detector, for which the nominal phase $\phi_{RF}^{0,T}$ on the scanning circle is selected arbitrarily, the time moment $t^i$ is transposed to total phase $\Phi_{RF}^i$; here the phase $\phi_{RF}^T(t)$ contains random and systematic deviations relative to the ideal periodic variations, due to the RF phototube. The corresponding time instability is given by

$$x_{RF}^T(t) = \frac{\phi_{RF}^T(t)}{2\pi v_{RF}^0}.$$

The technical time instability $x_{RF}^T(t)$ of the stand-alone streak camera tube is $\leq 10 \times 10^{-15}$/sec [10]. The technical time instability of the RF phototube is expected to be the same order of magnitude. The time moment $t^{i+1}$ for the next photoelectron will be transposed to

$$\Phi_{RF}^{i+1} = 2\pi v_{RF}^0 t^{i+1} + \phi_{RF}(t^{i+1}) + \phi_{RF}^T(t^{i+1}) + \phi_{RF}^{0,T}.$$

The time duration $\Delta t_{RF} = t^{i+1} - t^i$ is measured through total phase differences

$$\Delta\Phi_{RF} = \Phi_{RF}^{i+1} - \Phi_{RF}^i = 2\pi v_{RF}^0 (t^{i+1} - t^i) + \Delta\phi_{RF} + \Delta\phi_{RF}^T,$$

where $\Delta\phi_{RF} = \phi_{RF}(t^{i+1}) - \phi_{RF}(t^i)$, $\Delta\phi_{RF}^T = \phi_{RF}^T(t^{i+1}) - \phi_{RF}^T(t^i)$ and $v_{RF}^0$ is the nominal frequency of the RF oscillator. In this way the absolute value of the duration of two photoelectrons $t^{i+1} - t^i$ is expressed through the value of $v_{RF}^0$, with the random and systematic deviations relative to the ideal period equal to $\Delta x_{RF} + \Delta x_{RF}^T$, where $\Delta x_{RF} = \Delta\phi_{RF}/2\pi v_{RF}^0$, and $\Delta x_{RF}^T = \Delta\phi_{RF}^T/2\pi v_{RF}^0$. These figures are very small for the highly stable oscillators and for short ($t^{i+1} - t^i < 1$ sec) time intervals.

The CW photon beam produces a CW photoelectron flux, which forms a circle on the screen. The scanning speed of the PEs on the detector is $\upsilon = 2\pi R/T = 2\pi v_{RF}^0$. Here $T = 1/v_{RF}^0$ is the period of the RF field and $R$ is the radius of the circle on the detector. For example, if $T = 2 \times 10^{-9}$ s ($v_{RF}^0 = 500$ MHz) and $R = 2$ cm, we have $\upsilon \geq 0.5 \times 10^{10}$ cm/s.

The short photon pulse produces a short photoelectron pulse, which forms a spot somewhere on the scanning circle depending on the phase of the applied RF field. By measuring coordinates of the spot, the phase of the RF field and consequently the time can be determined. The SE detector provides fast nanosecond electrical signals like regular photomultipliers, which can be used for event by event processing of each PE. The technical (or reading) time resolution by definition is $\Delta\tau_d = d/\upsilon$, where $d$ is the size of the electron beam spot or the position resolution of the SE detector (if the electron spot is smaller). The time resolution $\Delta\tau_d \leq 20$ ps rms for a single photoelectron with $d = 1.0$ mm, $v_{RF}^0 = 500$ MHz and $R = 2$ cm.

Position of the photoelectrons on the circle can be determined by, e.g., using a direct readout scheme such as an array of small ($\sim 1 mm^2$) pixels with one readout channel per pixel. In this case the RF phototube operates as a 50 GHz sampling optoscope and can be used as an optical waveform digitization device in the nanosecond and subnanosecond domain with ~2 ps internal timing resolution.

### 3.3 Stand-alone operation: periodic photon source

Suppose the incident photon beam is the mode-locked femtosecond laser train with a nominal photon pulse frequency $v_{ph}^0 = 1/T_{ph}$ (see Fig.2). This OFC can be expressed mathematically as:

$$V_{ph}(t) = V_{ph}^0 \sin[2\pi v_{ph}^0 t + \phi_{ph}(t) + \phi_{ph}^0].$$

The quantity $\phi_{ph}(t)$ contains the deviations, both random and systematic, relative to the ideal periodic variations and $\phi_{ph}^0$ represents the nominal phase of photon train.

The RF phototube driven signal can be expressed as:

$$V_{RF}(t) = V_{RF}^0 \sin[2\pi v_{RF}^0 t + \phi_{RF}(t) + \phi_{RF}^0].$$

In such a case the RF phototube operates like a heterodyne for $V_{RF}(t)$ and $V_{ph}(t)$ signals. The result can be expressed as:

$$V_H(t) = V_H^0 \cos[2\pi(v_{RF}^0 - v_{ph}^0)t + \phi_{RF}^0 - \phi_{ph}^0 + \phi_{RF}(t) + \phi_{RF}^T(t) - \phi_{ph}(t)].$$

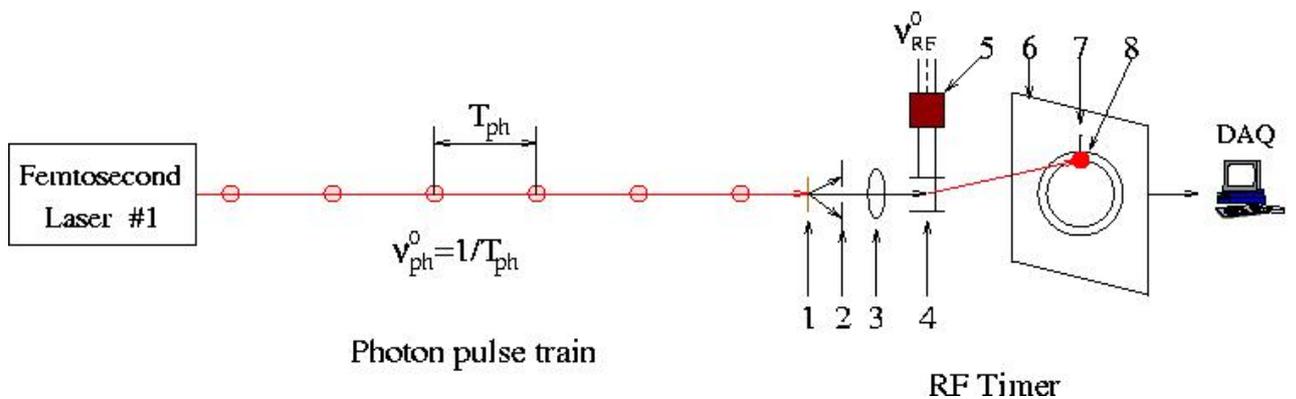

Figure 2: Schematic layout of the RF phototube with periodic photon source. 1- photo-cathode, 2- electron transparent electrode, 3- electrostatic lens, 4- RF deflector, 5- $\lambda/4$ coaxial RF cavity, 6- SE detector, 7- arbitrary reference, 8- image of PEs.

The quantity $\phi_{RF}^T(t)$ contains random and systematic deviations due to the RF phototube. The total phase of the heterodyned signal is:

$$\Phi_H(t) = 2\pi(\nu_{RF}^0 - \nu_{ph}^0)t + \phi_{RF}^{0,T} - \phi_{ph}^{0,T} + \phi_{RF}(t) + \phi_{RF}^T(t) - \phi_{ph}(t).$$

The PE beam spot will drift on the scanning circle with speed $\upsilon = 2\pi R(\nu_{RF}^0 - \nu_{ph}^0)$, where $R$ is the radius of the circle. The drift is clockwise for $\nu_{RF}^0 > \nu_{ph}^0$ and counterclockwise for $\nu_{RF}^0 < \nu_{ph}^0$. This feature can be used for precise comparison of two close frequencies, operational frequency of the RF deflector, $\nu_{RF}^0$, with the photon pulse train frequency $\nu_{ph}^0$. Continuous wave operation of the OFC and the RF phototube is the essential requirement for such a heterodyning system. The amplitude $V_H^0$ is determined by the RF phototube readout system. For time instability of the heterodyned signal we have:

$$\Delta x_H(t) = (1/2\pi\nu_H)\Delta\phi_H(t),$$

where $\nu_H = |\nu_{RF}^0 - \nu_{ph}^0|$ and $\Delta\phi_H(t) = \phi_{RF}(t) + \phi_{RF}^T(t) - \phi_{ph}(t)$.

The case when $\nu_{RF}^0 = \nu_{ph}^0$ is known as a synchroscan mode and ideally with no system drift, i.e. when $\Delta x_H(t) = 0$, the position of the PEs spot should stay stable on the scanning circle. In reality, the time instability of the heterodyned signal is determined by instabilities of the RF oscillator, the femtosecond photon train and the RF phototube. For precision measurements one needs to minimize these three ingredients. The $\phi_{RF}(t)$ and $\phi_{ph}(t)$ can be minimized by using a stable oscillator and mode-locked femtosecond laser train, while random and systematic deviations due to the RF phototube $\phi_{RF}^T(t)$ can be avoided by operating the tube in a synchroscan mode with a reference femtosecond photon train.

### 3.4 RF phototube in synchroscan mode

Figure 3 shows a schematic diagram of a standard synchroscan mode of an RF phototube.

The signal related to the incident photon pulse train with a repetition rate $\nu_{ph}^0$ can be expressed as:

$$V_{ph}(t) = V_{ph}^0 \sin[2\pi\nu_{ph}^0 t + \phi_{ph}(t) + \phi_{ph}^0].$$

This beam is split in two parts. The first one is directed to the photodiode and RF synthesizer, which generates a sinusoidal signal $V_{RF}(t)$ synchronous to the femtosecond photon train for driven of the RF phototube. This signal can be expressed as:

$$V_{RF}(t) = V_{RF}^0 \sin[2\pi\nu_{ph}^0 t + \phi_{ph}(t) + \phi_{RF}^T(t) + \phi_{ph}^{0,T}].$$

The second part of the photon beam illuminates the photocathode. The resulting photoelectrons are extracted, accelerated, focused, deflected by an RF deflector and detected with total phase:

$$\Phi_{RE}(t) = \phi_{RF}^T(t) + \phi_{RF}^{0,T} - \phi_{ph}^0.$$

The random and systematic deviations relative to the ideal period are determined only by the RF synthesizer and phototube and are equal to $\Delta x_{RF} = \Delta\phi_{RF}^T / 2\pi\nu_{ph}^0$. For the ideal tube with no drift, i.e. $\Delta x_{RF} = 0$, the position of photoelectrons on the scanning circle should stay constant, i.e. in synchroscan mode $\phi_{ph}(t)$ does not produce drift for photoelectrons from the same photon beam, even if it is not equal to 0.

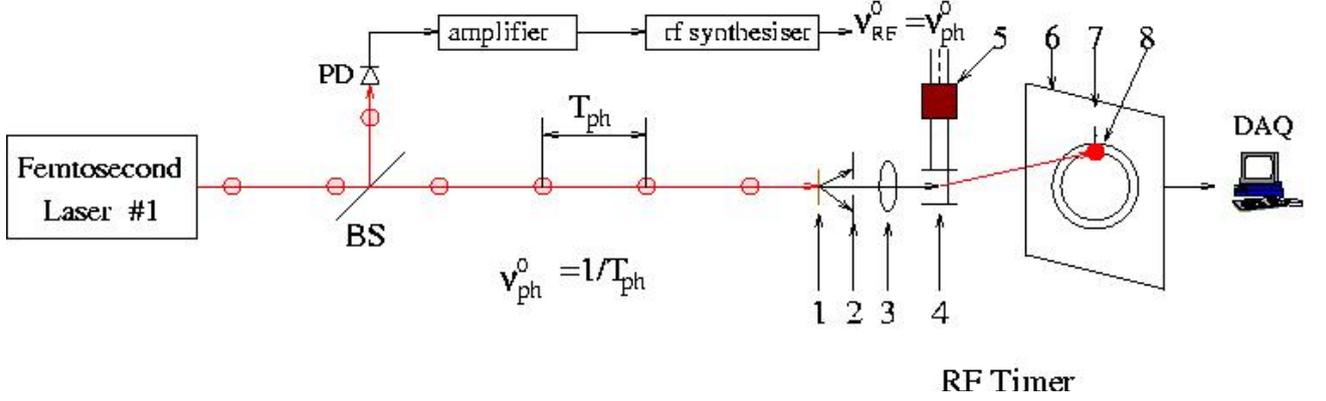

Figure 3: Schematic diagram of the experiment with RF phototube in the synchroscan mode. BS- beam splitter, PD- photo diode, 1- photocathode, 2- accelerating electrode, 3- electrostatic lens, 4- deflection electrode, 5- quarter wavelength coaxial cavity, 6- position sensitive secondary electron detector, 7- arbitrary reference fixed on SE detector readout system, 8- SEs from reference beam.

The temporal resolution of the technique is so high that every small component of instability in the RF synthesizer and phototube produces a noticeable drift. The drift phenomena of a standard synchroscan streak camera have been investigated in Ref. [10]. A measure of drift realized on two streak cameras at the same time and in the same conditions (by using the same synchroscan signal) shows that each camera has its own intrinsic and stochastic drift. The measured drift is linear against time and amounts of about 8 ps in 15 min (or less than 10 fs in 1 s) for the both streak cameras i.e. systematic error of relative time measurements performed by streak camera or RF phototube in short (< 1 s) term is less than 10 fs. In the case of long term measurements, systematic deviations due to $\phi_{RF}^{T}(t)$ can be avoided by using a reference photon beam. This method is schematically shown in Fig. 4. In this case the incident photon beam is split in two parts. The first one is directed to the photodiode and RF synthesizer, which generates a sinusoidal electrical signal $V_{RF}(t)$ synchronous to the photon train for driving the RF phototube. The second part of the photon beam illuminates the photocathode. The resulting photoelectrons are extracted, accelerated, focused, deflected by an RF deflector and detected. The position of PEs from the incident beam (8) on the SE detector (6) is related to an arbitrary fixed reference (7) on the position sensitive detector, and used as reference point.

The light pulses of the experimental investigations are also directed to the photocathode and detected simultaneously with the reference photon beam. The resulting photoelectrons are detected with total phase:

$$\Phi_{EXP}(t) = \phi_{RF}^{T}(t) + \phi_{RF}^{0,T} - \phi_{ph}^{0} + \phi_{EXP}.$$

Consequently, phase $\phi_{EXP}$ or time interval $\Delta T_{EXP}$ can be determined as

$$\phi_{EXP} = \Phi_{EXP} - \Phi_{REF},$$

and
$$\Delta T_{EXP} = \phi_{EXP} / 2\pi \nu_{ph}^0.$$

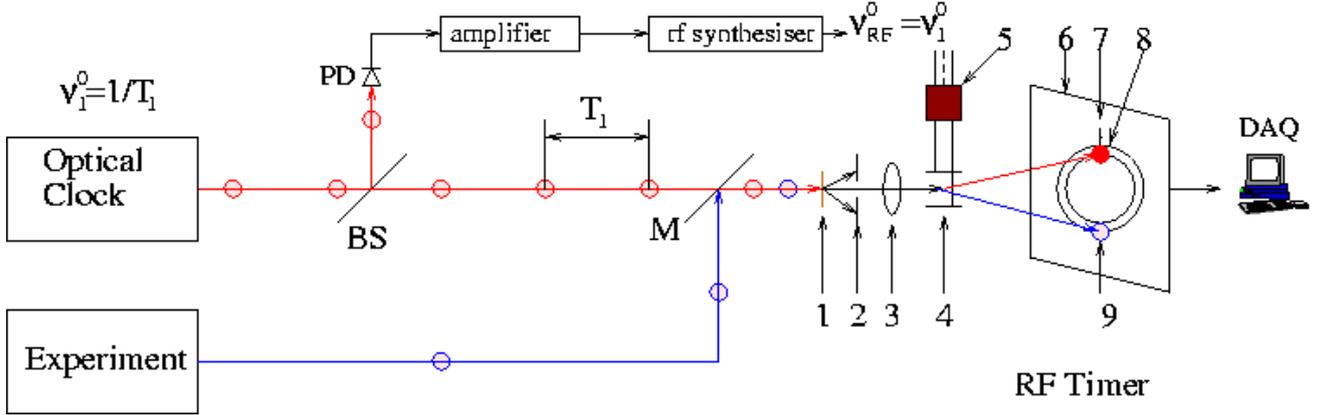

Figure 4: Schematic diagram of the experiment with reference beam. BS- beam splitter, PD- photo diode, 1- photocathode, 2- accelerating electrode, 3- electrostatic lens, 4- deflection electrode, 5- quarter wavelength coaxial cavity, 6- position sensitive secondary electron detector, 7- arbitrary reference fixed on SE detector readout system, 8- SEs from reference beam, 9- SEs from experiment.

The position of PEs from the experiment (9) is determined relative to the reference beam. In this way drifts related to the RF synthesizer and phototube are excluded and 200 fs instability for hours can be achieved [10]. The 200 fs is determined by statistics which achievable with streak cameras.

The RF phototube allows the performing of measurements with megahertz frequency and therefore has a powerful accumulating effect. Each measurement of the PE position with ~mm accuracy determines time with about $\Delta \tau_d \cong 20$ ps precision for the 500 MHz RF phototube. The centre of gravity of $N_1$ PEs is determined with precision $\Delta \tau_d / \sqrt{N_1}$. Sampling this measurement $N_2$ times decreased the timing error to $\Delta \tau_d / \sqrt{N_1 N_2}$ ps. By using reference beam with megahertz frequency ($N_2 = 10^6$) and power that provide $N_1 = 10^2$ PEs from the short ps bunches, the timing statistical error can be decreased to 2 fs in a 1 s. Consequently, the long term instabilities of the timing system due to systematic drifts can be decreased to 10 fs. This is a High Rate, High Resolution and Highly Stable $H^3$ timing technique.

It is interesting to note that in the case of regular timing techniques with regular PMTs and electronics, about 20 fs statistical precision can be reached while systematic timing drift, even with reference beam, lies in the ps level (see e.g. section 7.2.13 of Ref. [12]).

## 4. Possible applications: Precise measurements in nuclear physics
### 4.1 Precise time of flight system

We are proposing a new TOF system based on this new $H^3$ timing technique. The proposed system uses a pair of separated RF timers A and B, at distance *S*, to time the trip of a light pulses or elementary particles from one to the other. The setup is schematically displayed in Figure 5. Each RF timer consists of an RF phototube and RF oscillator, e.g., an optical clock and an RF synthesizer. Here we are using one optical clock for both RF timers. The OFC from the optical clock is split into three parts. One part is directed to a photodiode and generates a $\nu_0 \cong 500$ MHz sine signal, phase-locked with the photon train for driving the RF phototubes. The other two parts are directed to the left and right phototubes and serves as reference beams for the A and B RF timers. In this way the A and B RF timers are phase-locked with the optical clock. The time instability of such a system is determined by the time instability of the optical clock.

The experiment is to determine the TOF of light pulses or elementary particles from A to B. In such a time-of-flight experiment, the measurable quantities will be the phase differences of the RF timers A and B, measured relative to an arbitrarily chosen initial phase. In our case it means that an arbitrary calibration of the clocks will be used, e.g., by using arbitrarily calibrated position sensitive detectors of PEs. Here, the initial choice of phase difference, which is determined by distance $S$ between two clocks, RF frequency and arbitrarily calibrated position sensitive detectors of RF phototubes, amounts to a choice of synchronization; but once that choice is made, it is fixed once and for all. Experimentally this means measurement of the arc lengths on the arbitrary calibrated circles. For such a setup an absolute calibration can be established by using, e.g., light pulses or other reference particles. The light pulse way in Fig. 5 is shown separately.

Let us consider TOF measurements of the two types of particles: type 1 and type 2. Each event of such an experiment consists of four phases, $\Phi_1^A, \Phi_2^A, \Phi_1^B, \Phi_2^B$ and four corresponding times, $t_1^A, t_2^A, t_1^B, t_2^B$ measured relative to phases $\phi_{RF}^A(t_{ref}^A)$ and $\phi_{RF}^B(t_{RF}^B)$ of the A and B reference beams:

$$\Phi_1^A = 2\pi\nu_0 t_1^A + \phi_1^A + \phi_{RF}^A(t_1^A) + \phi_{OC}(t_1^A) - \phi_{RF}^A(t_{ref}^A) = 2\pi\upsilon_0 t_1^A + \phi_1^A + \phi_{OC}(t_1^A),$$

$$\Phi_2^A = 2\pi\nu_0 t_2^A + \phi_2^A + \phi_{RF}^A(t_2^A) + \phi_{OC}(t_2^A) - \phi_{RF}^A(t_{ref}^A) = 2\pi\upsilon_0 t_2^A + \phi_2^A + \phi_{OC}(t_2^A),$$

$$\Phi_1^B = 2\pi\nu_0 t_1^B + \phi_1^B + \phi_{RF}^B(t_1^B) + \phi_{OC}(t_1^B) - \phi_{RF}^B(t_{ref}^B) = 2\pi\upsilon_0 t_1^B + \phi_1^B + \phi_{OC}(t_1^B),$$

$$\Phi_2^B = 2\pi\nu_0 t_2^B + \phi_2^B + \phi_{RF}^B(t_2^B) + \phi_{OC}(t_2^B) - \phi_{RF}^B(t_{ref}^B) = 2\pi\nu_0 t_2^B + \phi_2^B + \phi_{OC}(t_2^B).$$

Here $\phi_{1,2}^A$ and $\phi_{1,2}^B$ are nominal phases of the type 1 and type 2 particles measured by A and B RF timers. They are determined by the geometry of the experiment and architecture of the RF timers. The quantities $\phi_{RF}^A(t), \phi_{RF}^B(t)$ contain the random and systematic deviations of the A and B RF timers. Periodic calibration by using a reference beam allows exclusion of these deviations. The quantity $\phi_{OC}(t)$ contains the random and systematic deviations of the optical clock.

The time intervals which type 1 and type 2 particles need to travel from A to B $\Delta t_1^{BA} = t_1^B - t_1^A$ and $\Delta t_2^{BA} = t_2^B - t_2^A$, are determined by relevant phase differences:

$$\Delta\Phi_1^{BA} = \Phi_1^B - \Phi_1^A = 2\pi\nu_0(t_1^B - t_1^A) + \phi_1^B - \phi_1^A + \phi_{OC}(t_1^B) - \phi_{OC}(t_1^A);$$

$$\Delta\Phi_2^{BA} = \Phi_2^B - \Phi_2^A = 2\pi\nu_0(t_2^B - t_2^A) + \phi_2^B - \phi_2^A + \phi_{OC}(t_2^B) - \phi_{OC}(t_2^A).$$

Here $\phi_1^B - \phi_1^A = \phi_2^B - \phi_2^A$, and the figures of $\phi_{OC}(t_i^B) - \phi(t_i^A)$ are negligibly small for the stable clock and for shot ($t_i^B - t_i^A < 1$ s) time intervals. The difference of the time intervals that type 1 and type 2 particles travel from A to B is

$$\Delta\Phi_2^{BA} - \Delta\Phi_1^{BA} = 2\pi\nu_0(\Delta t_2^{BA} - \Delta t_1^{BA}) = 2\pi\upsilon_0 \delta\tau.$$

The time difference $\delta\tau = \Delta t_2^{BA} - \Delta t_1^{BA}$ is free from nominal phases and its error is determined mainly by statistic error of phase measurements.

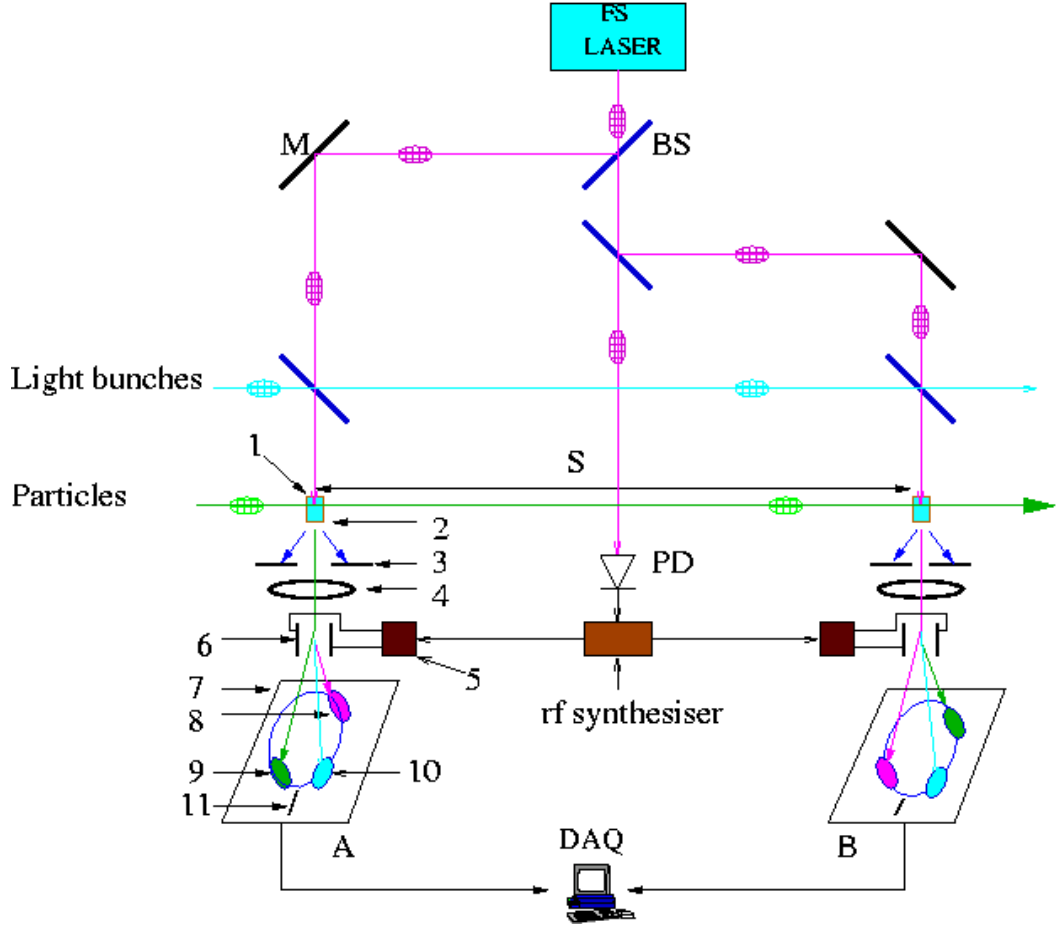

Figure 5: Schematic of the TOF system. BS- beam splitter, M- reflector, PD- photo diode, 1- Cherenkov radiator or scintillator, 2- photocatode, 3- accelerating electrode, 4- electrostatic lens, 5- quarter wavelength coaxial cavity, 6- deflection electrode, 7- position sensitive secondary electron detector, 8- reference PEs, 9 and 10- PEs from elementary particles and light bunches, respectively, 11- arbitrary reference fixed on SE detector readout system.

Let us consider the currently available parameters: a femtosecond laser beam, e.g., a continuous photon beam with a repetition rate of about 1 MHz and pulse lengths $\leq 1$ ps. The relative precision of phase measurement for a single PE is about $10^{-2}$. For a 500 MHz RF deflector this corresponds to 20 ps. It is assumed that each photon pulse or particle produces about 100 PEs, the center of gravity of which is determined with ~2 ps precision. The charged particles will be timed by digitization of Cherenkov photons [12] or fast scintillation light (see Ref. [13] and next paragraph) with ~10 ps precision or better. Consequently, each measurement will provide the values of $\Delta t_1^{BA}$ and $\Delta t_2^{BA}$ with precision of about ~3 ps and ~15 ps, respectively. Their difference $\delta\tau = \Delta t_2^{BA} - \Delta t_1^{BA}$ will be determined with ~20 ps precision or better. Sampling this measurement $N = 10^6$ times the timing error decreased to ~20 fs.

The proposed technique is a nearly ideal tool for TOF measurement because the precision is determined by statistics only.

### 4.2 Precise timing by using fast scintillators with RF phototube

When a charged particle passes through the scintillator bar, it excites molecular levels of the scintillator which emit bands of UV light. The wavelength shifter transforms UV light into the blue wavelength region detectable by photocathodes. We propose a precision time measuring system based on first arriving photons, detected by means of the RF phototube.

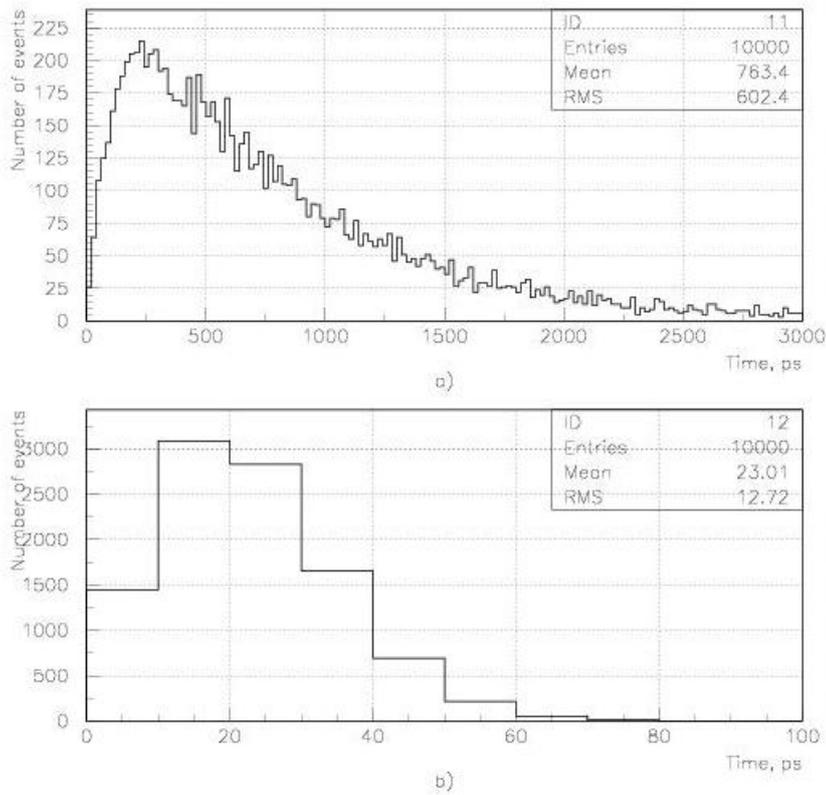

Figure 6: a) - two exponential fluorescence decay time distribution; b) - early arriving photoelectron time distribution from total number of 250 photoelectrons.

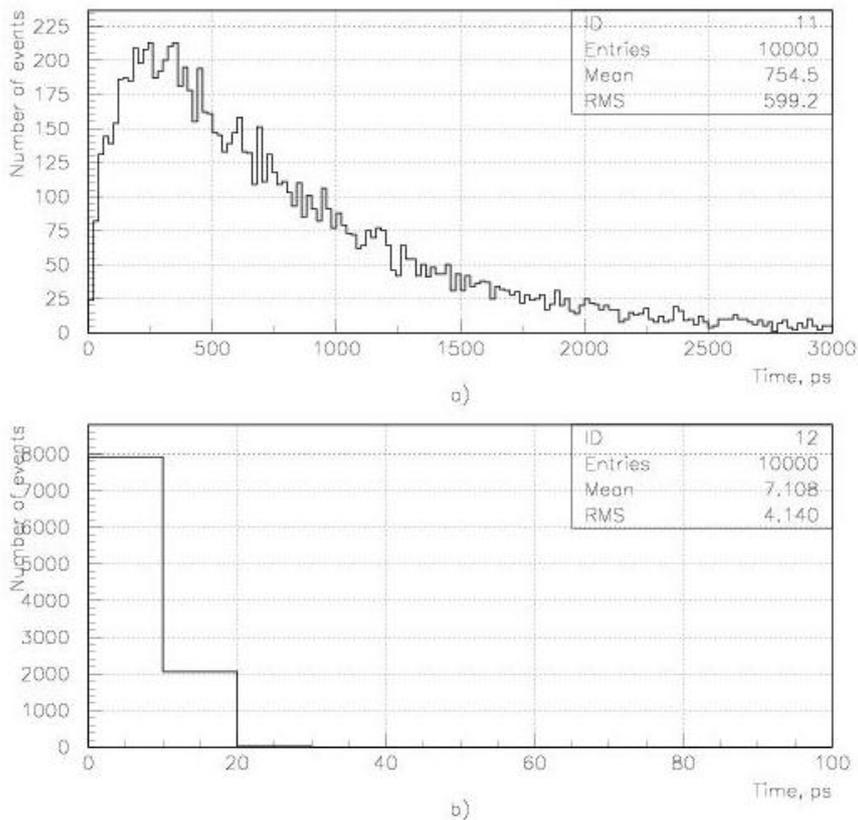

Figure 7: a) - two exponential fluorescence decay time distribution; b) - early arriving photoelectron time distribution from total number of 2500 photoelectrons.

The characteristics of such a system were investigated by using Monte Carlo (MC) simulations. The following dominant factors have been taken into account in these simulations:
- The fluorescence time distribution: two exponential time distribution have been used with 100 ps and 700 ps rise and decay times, typical for ultrafast BC-422Q scintillators.
- The timing accuracy of the photon detector: 20 ps was used.

The distributions of two exponential fluorescence decay times and first photoelectron times from total number of photoelectrons equal to 250 and 2500 are shown in Fig. 6[a,b] and Fig. 7[a,b] respectively. The MC simulations shows that the time resolution, rms, of such a time correlated scintillator counter with RF phototube is about 10 ps or better.

### 4.3 Absolute calibration of magnetic spectrometer by TOF measurement

The key point of the PR-08-012 experiment is determination of the binding energies of light hypernuclei by decay $\pi^-$ momentum measurement. These momentums lay in the range around 100 MeV/c and it was proposed to use the high resolution magnetic spectrometer $H\pi S$ for their measurement. It seems that this technique is the most proper one for this experiment. However the quest of absolute calibration of the magnetic spectrometer within $10^{-4}$ precision is remains open. We propose to use the high resolution, high rate and highly stable TOF system located on the focal plane of the $H\pi S$ for absolute calibration of the magnetic spectrometer by TOF measurement. The fundamentals of the TOF can be summarized as follows. The TOF of particle with mass $m_1$ and momentum $p$ is for a flight path $L$

$$\Delta t_1 = L/(\beta_1 c) = (L/c)\sqrt{1 + m_1^2 c^2 / p^2} \ ,$$

where $c$ is the speed of light. The difference of TOF of two particles with mass $m_1$ and $m_2$ and momentum $p$ is:

$$\Delta t_{12} = (L/c)(\sqrt{1 + m_1^2 c^2 / p^2} - \sqrt{1 + m_2^2 c^2 / p^2}).$$

The difference of TOF of particle with mass $m_1$ and light photons is

$$\Delta t_{1L} = (L/c)(\sqrt{1 + m_1^2 c^2 / p^2} - 1).$$

From these relations it follows that precise measurement of the TOF or the difference of TOF of particle pairs, for a fixed momentum $p$ and for a known flight path $L$, allows determination of the absolute value of $p$. Indeed from the TOF measurement we have

$$p = m_1 c (L/c) / \sqrt{(\Delta t_1)^2 - (L/c)^2} \ .$$

The magnetic rigidity for the relativistic charged particle can be expressed as

$$B\rho = \frac{p}{q} = \frac{\gamma m v}{q} = \frac{m \gamma \beta c}{q} \ ,$$

where $v$, $m$ and $q$ are the velocity, rest mass and charge of the particle and relativity variables are determined as

$$\beta = v/c \ , \ \gamma = \frac{1}{\sqrt{1 - \beta^2}} \ , \ \gamma\beta = \frac{\beta}{\sqrt{1 - \beta^2}} \ .$$

The relative differential is

$$\frac{d(B\rho)}{B\rho} = \frac{dp}{p} = \frac{d(\gamma\beta)}{\gamma\beta} \ .$$

The differential of $\beta\gamma$ is

$$d(\gamma\beta) = \frac{\partial(\beta/\sqrt{1-\beta^2})}{\partial\beta}d\beta = (\frac{1}{\sqrt{1-\beta^2}} - \frac{\beta^2}{(1-\beta^2)^{3/2}})d\beta = (\gamma + \beta^2\gamma^3)d\beta = \gamma(1+(\beta\gamma)^2)d\beta,$$

and the relative expression is

$$\frac{d(\gamma\beta)}{(\gamma\beta)} = (1+(\gamma\beta)^2)\frac{d\beta}{\beta} = \gamma^2\frac{d\beta}{\beta},$$

where

$$1+(\gamma\beta)^2 = 1 + \frac{\beta^2}{1-\beta^2} = \frac{1-\beta^2+\beta^2}{1-\beta^2} = \gamma^2 \text{ was used.}$$

Therefore

$$\frac{d(B\rho)}{B\rho} = \frac{dp}{p} = \frac{d(\gamma\beta)}{(\gamma\beta)} = \gamma^2\frac{d\beta}{\beta} = \gamma^2\frac{dt}{\Delta t_1},$$

where $\frac{dt}{\Delta t_1}$ is the relative error of the TOF measurement. For $p_1 = 100$ MeV/$c$ and $L = 1$ m, $\Delta t_1 = 5.727$ ns. In real life, determination of the absolute value of momentum means selecting events in small momentum range, e.g. in the $100 \pm 0.01$ MeV/$c$ range, by use of a high resolution magnetic spectrometer, and measuring $\Delta t_1$ for these events. Each measurement of the $\Delta t_1$, with 20 ps accuracy, determines absolute momentum with about 0.53 MeV/$c$ precision. Sampling this measurement $N$ times decreased the errors to $0.02/\sqrt{N}$ ns and $0.53/\sqrt{N}$ MeV/$c$ respectively, which for $N = 10^4$ amounts to 200 fs and 5.3 keV/$c$. The Monte Carlo distributions of $\Delta t_1$ and $p$ are shown in Fig. 8 (a) and (b), respectively.

In reality, precise measurements of the path length and absolute measurements of magnetic rigidity are very difficult. Nevertheless, as the path length and the magnetic rigidity can be considered to be constant with in relative precision better than $10^{-4}$ for a fixed configuration of the system, the direct relation between the time-of-flight and the momentum and/or magnetic rigidity can be found using the electrons for determination of the $L/c$ of the TOF system. This is because for 100 MeV/c electrons $1-\beta < 10^{-4}$.

## 5. Summary
The crucial point of the JLab Experiment PR-08-012 is absolute calibration of the high resolution pion spectrometer $H\pi S$ with accuracy $10^{-4}$ or better. We are proposing a new TOF system based on an RF phototube and optical clock. It is demonstrated that the proposed TOF system can achieve 20 fs accuracy level (to be compared to ~1 ps, available with the regular timing technique) and enables absolute calibration of the $H\pi S$ for ~100 MeV/c pions within accuracy $10^{-4}$ - $10^{-5}$.

## 6. Acknowledgement
Author thanks to S. Corneliussen for assistance.

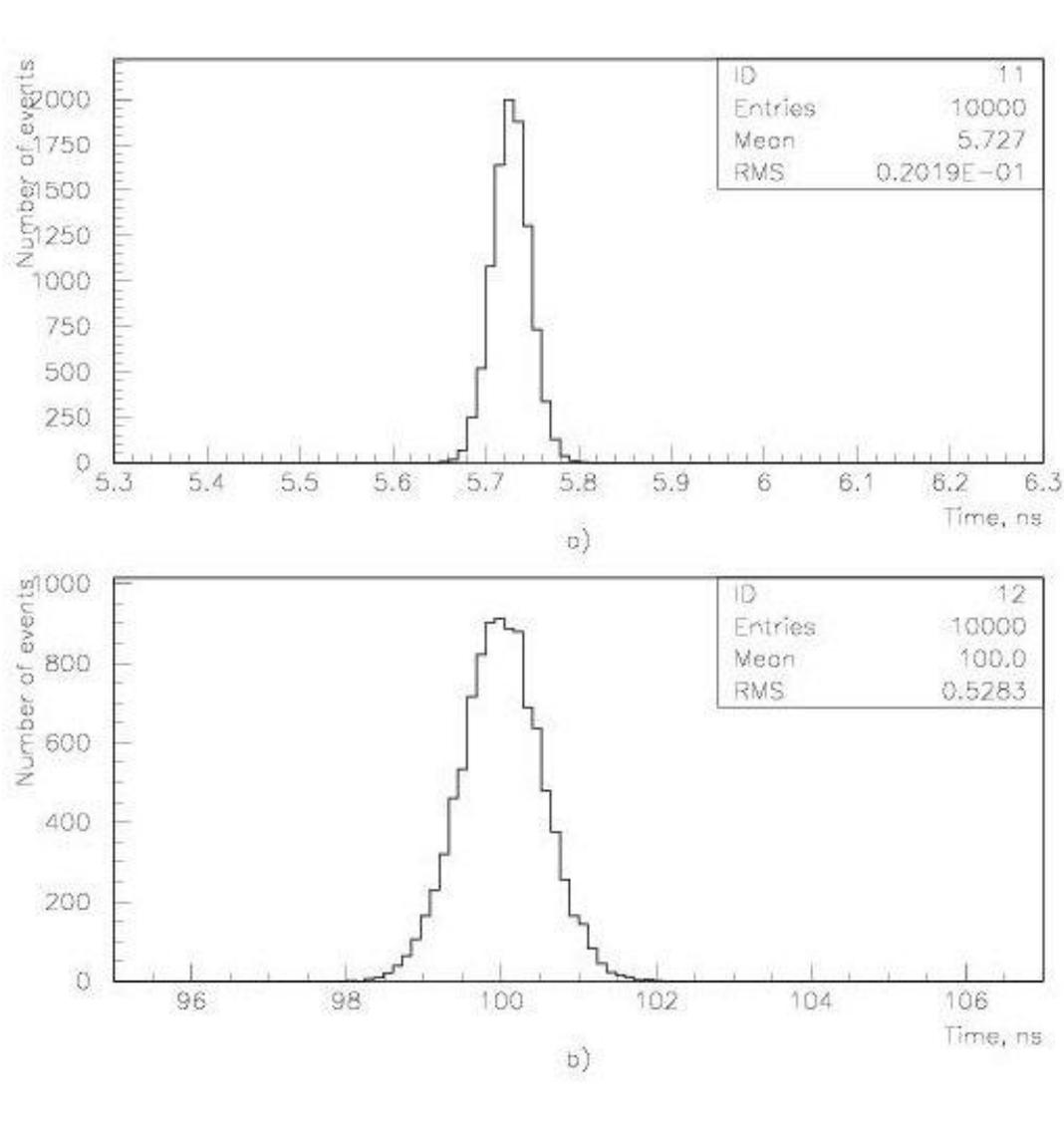

Figure 8: Distributions of time-of-flight (a) of (a) and reconstructed momentum (b) of p = 100 MeV/c pions. Number of events is $N = 10^4$ and the average momentum can be determined within error 5.3 keV/c.